\documentclass[aps,prb,twocolumn,showpacs,superscriptaddress]{revtex4-1}
\usepackage{graphicx,bm,amsmath,dcolumn,amssymb,color}
\graphicspath{{figures}}
\usepackage{longtable,geometry,hyperref}

\begin{document}

\title{Sublattice extraordinary-log phase and new special point of the antiferromagnetic Potts model}
\author{Li-Ru Zhang}
\affiliation{School of Microelectronics $\&$ Data Science, Anhui University of Technology, Maanshan, Anhui 243002, China}

\author{Chengxiang Ding}
\email{dingcx@ahut.edu.cn}
\affiliation{School of Microelectronics $\&$ Data Science, Anhui University of Technology, Maanshan, Anhui 243002, China}

\author{Wanzhou Zhang}
\affiliation{College of Physics and Optoelectronics, Taiyuan University of Technology, Shanxi 030024, China}



\author{Long Zhang}
\email{longzhang@ucas.ac.cn}
\affiliation{Kavli Institute for Theoretical Sciences and CAS Center for Excellence in Topological Quantum Computation, University of Chinese Academy of Sciences, Beijing 100190, China}

\date{\today}

\begin{abstract}
We study the surface criticality of a three-dimensional classical antiferromagnetic Potts model, whose bulk critical behaviors belongs to the XY model because of emergent O(2) symmetry.  We find that  the surface antiferromagnetic next-nearest neighboring interactions can drive the extraordinary-log phase to the ordinary phase, the transition between the two phases belongs to the universality class of the well-known special transition of the XY model. Further strengthening the surface next-nearest neighboring interactions, the extraordinary-log phase reappears,  but the main critical behaviors are dominated on the sublattices of the model;  the special point between the ordinary phase and the sublattice extraordinary-log phase belongs to a  new universality class. 

\end{abstract}

\maketitle

\section{Introduction}
For a system at the critical point of a phase transition, the correlation function and order parameter and some other physical quantities exhibit power-law scaling behaviors, which are called critical behaviors.  It is a hot topic in statistical physics
with a long history of research.   

At the critical point, the critical behaviors not only manifest in the bulk but also on the surface, which is called the surface critical behaviors\cite{binder1983}. Depending on the strength of the surface interactions, the surface critical behaviors can be richer than the bulk critical behaviors.  Generally, there are three types of surface critical phases, dubbed `ordinary phase', 	`special point', and `extraordinary phase'. Ordinary phase refers to the case when the surface interactions are not too strong, and the surface critical behavior are purely induced by the bulk critical state; extraordianry phase refers to the case when the surface interactions are strong enough,  the surface has become ordered or critical with logarithmic scaling behaviors (extraordinary-log phase); the special point is the transition point between the ordinary phase and the extraordinary phase.  Typical examples can be found in the classical O($n$) spin models\cite{On,O4,O3sp,XYlog}. 

The research of surface critical behaviors also has long history, and the interest in this field are renewed by the recent works in quantum spin models\cite{Long2017, Ding2018}. A series of  related studies greatly promote the interest  in this field\cite{Weber2018,Weber2019,Max2022,Zhu2021,Weber2021,Jian2021,Toldin2021,Ding2021,Yu2021,Zhu2021-2,Max2022a, Ding2022,clock, Lv2022a,Lv2022b}, which includes the finding of the extraordinry-log phase\cite{Max2022}, a great breakthrough in the research of critical phenomena.  Such type of critical phase, characterized by the logarithmic decaying of correlation function, has already been numerically found in both the classical spin models\cite{O3sp,XYlog,Ding2022} and also the quantum spin models\cite{Lv2022a,Lv2022b}.  

In our recent work\cite{Ding2022}, we find the extraordinary-log phase in the classical antiferromagnetic Potts model, which has a bulk critical point belongs to the XY model because of the emergent O(2) symmery\cite{Ding2016}, although the spin symmetry of the model is discrete.  Tunning the surface nearest neighboring (NN) interactions lead to a surface phase diagram similar to that of the XY model. 
More importantly, by adding ferromagnetic next-nerearest neighboring(NNN)  interactions to the surface, a phase transition from the extraordinary-log phase to the ordered phase is found, whose critical behaviors are very different from the traditional ordered-disordered phase transition. 
In the current paper, with the purpose of exploring new surface critical behaviors,  we add {\it antiferromagnetic}  NNN interactions to the surface, we find that the  antiferromagnetic NNN interactions can drive the surface from the extraordinary-log phase to an ordinary phase, and then drive the system to a new extraordinary-log phase whose main critical behaviors are dominated on the sublattices if the strength of the NNN interactions are strong enough.
The universality class of the special point between the ordianry phase and the sublattice extraordinary-log phase is different from the well-known special point of the XY model. 

\begin{figure}[htpb]
\includegraphics[width=0.9\columnwidth]{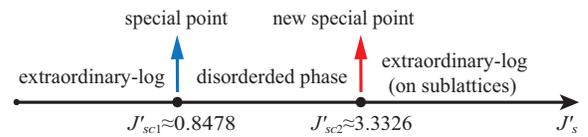}
\caption{Surface phase diagram of the antiferromagnetic Potts model (\ref{Ham})with $T=T_c^{\rm bulk}=1.22603$ and  $J_{\rm s}=5$.}
\label{PD}
\end{figure}

The phase diagram of the model is shown in Fig. \ref{PD} and the paper is organized as follows: In Sec. \ref{model}, we introduce the model and the method;  in Sec. \ref{results}, we present the numerical results, which include the critical properties of the two special points, the ordinary phase, and the new extraordinary-log phase.  We conclude our paper in Sec. \ref{conclusion}.

\section{Model and Method}
\label{model}
The Hamiltonian for the three-state Potts model with surface NNN antiferromagnetic interactions on a simple cubic lattice is defined as
\begin{eqnarray}
	\mathcal{H}=J\sum\limits_{\langle i,j\rangle}\delta_{\sigma_i,\sigma_j}
	+J_s\sum\limits_{{\langle i,j\rangle}^s}\delta_{\sigma_i,\sigma_j}
	+J_s^\prime\sum\limits_{{\langle\langle i,j\rangle\rangle}^\prime}\delta_{\sigma_i,\sigma_j},\label{Ham}
\end{eqnarray}
where $\langle i,j\rangle$, ${\langle i,j\rangle}^s$ and ${\langle\langle i,j\rangle\rangle}^\prime$ denote the bulk NN, the surface NN, and the surface NNN sites, respectively. All the interactions are antiferromagnetic, and the spin $\sigma_i$ can be mapped to unit vectors on the plane:
\begin{eqnarray}
\vec{\sigma}_i=(\cos\theta_i,\sin\theta_i),
\end{eqnarray}
with $\theta_i=2\pi\sigma_i/3$ and $\sigma_i=1,2,3$. The bulk phase transition of the model belongs to the XY universality because of emergent O(2) symmetry\cite{Ding2016}, and the critical point has been refined by extensive Monte Carlo simulations in Ref. \onlinecite{Ding2022}, which is $T_c=1.22603$.  When there is no NNN interaction and the strength of the  NN interactions $J_s>2.04119$, the surface is in the extraordinary-log phase\cite{Ding2022}. 

For simulations of the model, we adopt a combination of the local update (Metropolis algorithm) and the  geometric clustering algorithm\cite{geo}, making us able to study the system with size up to $L=128$. 
The surface variables we sampled include the surface squared (staggered) magnetization $m_{\rm s1}^2$ and the surface magnetic susceptibility $\chi_{\rm s1}$,which are defined as,
\begin{eqnarray}
m_{\rm s1}^2=\langle \mathcal{M}_{\rm s1}^2 \rangle, \label{ms12}
\end{eqnarray}
\begin{eqnarray}
\chi_{\rm s1}=L^2(\langle\mathcal{M}_{\rm s1}^2\rangle-\langle|\mathcal{M}_{\rm s1}|\rangle^2),
\end{eqnarray}
where $\mathcal{M}_{\rm s1}$ is defined as,
\begin{eqnarray}
\mathcal{M}_{\rm s1}&=&\frac{1}{L^2}\sum\limits_{\vec{R}}(-1)^{x+y+z} \vec{\sigma}_{\vec{R}} \label{Ms}.
\end{eqnarray}
Here $\vec{R}=(x,y,z)$ is the coordination, for the surface, $z$ should be 1 or $L$, and $L^2$ is the number of sites of the surface. We also sample the sublattice squared (staggered)  magnetization $m^2_{\rm s1A}$, which are defined similar to (\ref{ms12}), but the sites are restricted to be in the  sublattice A of the surface. (The square surface is bipartite, which can be divided into two equivalent sublattices) 

We also sample the surface specific heat $C_{\rm v1}$,
\begin{eqnarray}
C_{\rm v1}&=&L^2(\langle \mathcal{E}_{\rm 1}^2\rangle-\langle \mathcal{E}_{\rm 1}\rangle^2)/T^2,\label{Cv}
\end{eqnarray}
where $T$ is the temperature, $\mathcal{E}_{\rm 1}$ is the microscopic energy density of the surface.

The surface correlation function is defined as
\begin{eqnarray}
C_\parallel(r)=\langle\vec{\sigma}_i\cdot\vec{\sigma}_{i+r}\rangle, \label{Cr}
\end{eqnarray}
where the site $i$ and  $i+r$ are restricted to be on the surface.
The surface correlation length  $\xi_{\rm 1}$ and ``structure factor" $F_1$ are defined as
\begin{eqnarray}
&&\xi_{\rm 1}=\frac{(m^2_{\rm s1}/F_{\rm1 }-1)^{1/2}}{2\sqrt{\sum\limits_{i=1}^d \sin ^2(\frac{k_i}{2})}} \; ,\\
&&F_{\rm 1}=\frac{1}{L^4}\Big\langle\big|\sum\limits_{\vec{R}}(-1)^{x+y+z}e^{i\vec{k} \cdot\vec{R}}\vec{\sigma}_{\vec{R}}\big|^2\Big\rangle  \label{F}
\end{eqnarray}
where $\vec{k}$ is the ``smallest wavevector" along the $x$ direction--i.e., $\vec{k} \equiv (2 \pi/L, 0)$.
In the disordered phase, the correlation length $\xi_1$ is finite and the correlation ratio $\xi_1/L$ decreases to zero, while in the ordered phase or extraordinary-log phase $\xi_1/L$ diverges rapidly due to the rapid disappearance of the ``structure factor" $F_1$.
The correlation ratio $\xi_1/L$ in the critical phase has a finite nonzero value in the thermodynamic limit.
Therefore, the correlation ratio $\xi_1/L$ is a good tool for locating the critical point of the phase transition.

\section{Results}
\label{results}

\subsection{Two special points}
\begin{figure}[htpb]
\includegraphics[width=1.0\columnwidth]{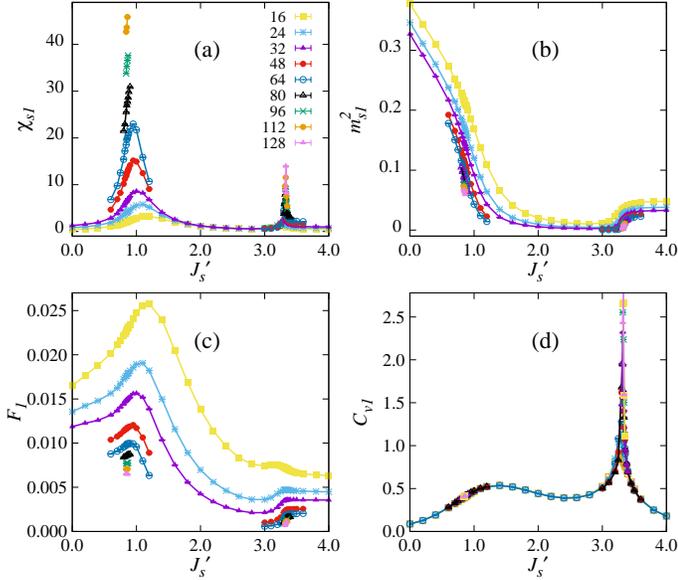}
\caption{Surface critical behavior of the antiferromagnetic Potts model (\ref{Ham}) at $T=T_c^{\rm bulk}=1.22603$, $J_s=5$: (a)  surface susceptibility $\chi_{\rm s1}$, (b) surface squared magnetization $m^2_{\rm s1}$, (c) surface structure factor $F_1$, and (d) surface specific heat $C_{\rm v1}$.}
\label{psurf4}
\end{figure}

By the Monte Carlo method we simulate the antiferromagnetic Potts model (\ref{Ham}) with open boundary along the $z$-direction and periodic boundaries in the $x$- and $y$-  directions. 
At the bulk critical point $T=T_c^{\rm bulk}=1.22603$\cite{Ding2022}, setting the system surface initially to the extraordinary-log phase ($J_s=5$), varying the surface NNN interactions $J^\prime_s$, 
we can find two special points.
As shown in Fig. \ref{psurf4}(a), the surface magnetic susceptibility $\chi_{s1}$ of the system shows two peaks  at $J^{\prime(1)}_{\rm sc}\approx 0.85$ and $J^{\prime(2)}_{\rm sc}\approx3.3$, where the former peak diverges faster. 
The behaviors of the surface squared magnetization $m_{\rm s1}^2$ and the surface structure factor $F_1$, as shown in Figs. \ref{psurf4}(b) and (c),  also demonstrate the two transitions.
We can see that the  $m_{\rm s1}^2$  tends to zero in the region between the two critical points even for small system sizes, the system is in a disordered phase, which is different from the well known ``ordinary phase";  we will discuss the properties  of  such phase specifically in the next subsection.
The specific heat $C_{\rm v1}$, as shown in Fig. \ref{psurf4}(d),  demonstrates a rapidly diverging peak at the second phase transition point, however, it does not show any obvious peak for the first transition.

\begin{figure}[htpb]
	\includegraphics[width=0.9\columnwidth]{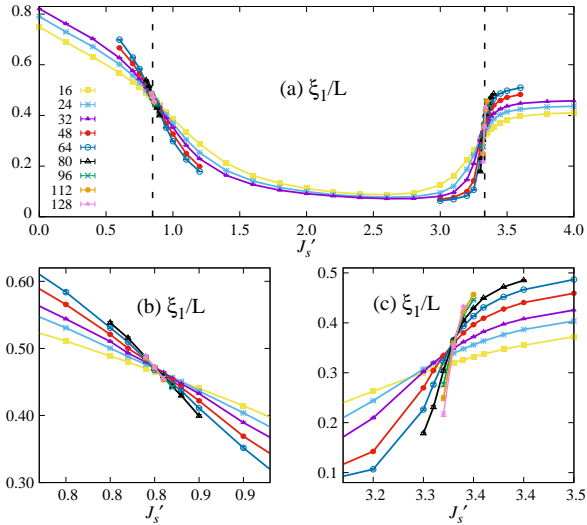}
	\caption{Surface correlation ratio $\xi_1/L$ of the antiferromagnetic Potts model (\ref{Ham}) with $T=T_c^{\rm bulk}=1.22603$ and  $J_{\rm s}=5$, the dashed line positions in (a) are $J^{\prime (1)}_{\rm sc}=0.8478$ and $J^{\prime (2)}_{\rm sc}=3.3326$; (b) and (c) are  the local enlargements of the vicinity of the two critical points.}
	\label{xi/L}
\end{figure}

In order to quantitatively determine the two critical points, we investigate the surface correlation ratio $\xi_1/L$, as shown in Fig. \ref{xi/L}, which obviously shows the two critical points. In the vicinity of the critical point, $\xi_1/L$  satisfies the  finite-size scaling (FSS) formula:
\begin{eqnarray}
	\xi_1/L=a_0+\sum\limits_{k=1}^{k_{\rm max}}a_k(J^\prime_{\rm s}-J^\prime_{\rm sc})^kL^{ky_s}+bL^{y_1},  \label{xi/LFSS}
\end{eqnarray}
where $y_s>0$ is the critical exponent, $J^\prime_{\rm sc}$ is the critical point , 
$y_1<0$ is the correction-to-scaling exponent, $a_0$, $a_k$, and $b$ are unknown parameters.

\begin{figure}[htpb]
\includegraphics[width=1.0\columnwidth]{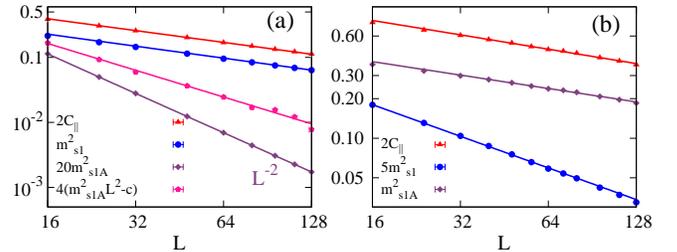}
\caption{(a)) log-log plot of $C_{\parallel}(L/2)$, $m^2_{\rm s1}$, $m^2_{\rm s1A}$,  and  $m^2_{\rm s1A}L^2-c$ versus $L$ at the first transition point $J^{\prime (1)}_{\rm sc}=0.8478$; (b)log-log plot of $C_{\parallel}(L/2)$, $m^2_{\rm s1}$ and $m^2_{\rm s1A}$ at the second transition point $J^{\prime (2)}_{\rm sc}=3.3326$.}
\label{jsc1c2}
\end{figure}

For the first transition, the data fitting of $\xi_1/L$ according to (\ref{xi/LFSS}) with $y_1=-1$ gives $J^{\prime(1)}_{\rm sc}=0.8478(9)$ and $y_{\rm s}=0.59(4)$. We can see that the critical exponent $y_s$ coincides with the result of the well-known special transition of the XY model\cite{On}; in order to confirm such transition is in the same universality of the XY model, we study the scaling behaviors of the squared magnetization $m_{\rm s1}^2$ and the correlation function $C_\parallel(L/2) $, at the critical point, they satisfy the following FSS formulas  
 \begin{eqnarray}
	&&m^2_{\rm s1}L^2=c+L^{2y_{\rm h1}-2}(a+bL^{y_1}) \label{msurf_stag}\\
	&&C_\parallel(L/2)=L^{-1-\eta_\parallel}(a+bL^{y_1}) \label{Corp}
\end{eqnarray}
where $y_{\rm h1}$ and $\eta_\parallel$ are the critical exponents. 
The data fitting, with  $y_1=-1$, gives  $y_{\rm h1}=1.698(4)$ and $\eta_\parallel=-0.399(5)$.  We can see that the values of $y_{\rm h1}$ and $\eta_\parallel$ satisfy the scaling formula\cite{sc1,sc2}
\begin{eqnarray}
	\eta_\parallel=d-2y_{h1}, \label{SL}
\end{eqnarray}
where $d=3$ is the space dimension of the system. This means that 
$m_{s1}^2$ and $C_\parallel$ should decay with the same power as the increasing of the system size, which is demonstrated in Fig.  \ref{jsc1c2}(a).
We can also see that the values of $y_{\rm h1}$ and $\eta_\parallel$ coincide with those of the well-known special point  of the XY model\cite{On}, i.e., they are in the same universality class.

We also investigate the scaling behaviors of the sublattice squared magentizaton $m^2_{\rm s1A}$, which is the same as $m^2_{\rm s1}$, specifically, $m^2_{\rm s1A}$ satisfies
 \begin{eqnarray}
	m^2_{\rm s1A}L^2=c+L^{2y_{\rm h1A}-2}(a+bL^{y_1}), \label{msA}
\end{eqnarray}
where we have written the exponent as $y_{\rm h1A}$.
The data of $m^s_{\rm s1A}$  is also shown in Fig. \ref{jsc1c2}(a).  We can see that $m^2_{s1A}$ shows a perfect power-lay scaling,  however, this is misleading, such power-law scaling comes from the nonsingular part, which is  the first term of the RHS of Eq. (\ref{msA}). Because  in the current case  $y_{h1A}<1$,  $m_{\rm s1A}^2$ is dominated by the term $c/L^2$. This is also the reason why we do not write the scaling form as $m^2_{\rm s1A}=c/L^2+L^{2y_{\rm h1A}-4}(a+bL^{y_1})$. 
The log-log plot of $m^s_{\rm s1A}L^2-c$ versus $L$ is also included in Fig. \ref{jsc1c2}(a), 
and the data fitting  according to Eq. (\ref{msA}) gives $y_{\rm h1A}=0.26(3)$.

For the second special  point, the data fitting gives  $J^{\prime(2)}_{\rm sc}=3.3326(8)$ and $y_{\rm s}=1.44(4)$. We can see that the critical exponent $y_{\rm s}$ is obviously different from the first one; in order to determine the universality class of this transition, 
we investigate the scaling behaviors of  the squared magnetization $m_{\rm s1}^2$ and the correlation function $C_\parallel(L/2) $; at the critical point, they satisfy the FSS formulas   (\ref{msurf_stag}) and (\ref{Corp}). 
By the data fitting, we get  $y_{\rm h1}=1.56(2)$ and $\eta_ \parallel=-0.59(3)$; 
these are  obviously different from the results of the first transition. Furthermore, we can see that they do not satisfy the scaling law (\ref{SL}),
 i.e., $m_{\rm s1}^2$  and $C_\parallel(L/2)$ decay with different powers, which are demonstrated in Fig. \ref{jsc1c2}(b).
In order to understand such discrepancy and the nature of this phase transition, we investigate the sublattice  squared (staggered)  magnetization $m^2_{\rm s1A}$, which is also shown in  Fig. \ref{jsc1c2}(b). The data fitting are performed in a similar way  according to Eq. (\ref{msA}),  which gives 
$y_{\rm h1A}=1.80(2)$. 
We can see that  $\eta_\parallel$ and $y_{\rm h1A}$ satisfy the scaling law (\ref{SL}), i.e., $C_{\parallel}$  and $m_{\rm s1A}^2$  decay with the same power, as shown in Fig. \ref{jsc1c2}(b). 
This result reveals the sublattice nature of such transition.
In summary, the second  transition is characterized by the following critical exponents
\begin{eqnarray}
	y_{\rm s}&=&1.44(4),\\
	y_{\rm h1}&=&1.56(2),\\
	y_{\rm h1A}&=&1.80(2),\\
	\eta_ {\parallel}&=&-0.59(3).
\end{eqnarray}

\subsection{The ordinary phase}
\begin{figure}[htpb]
\includegraphics[width=1.0\columnwidth]{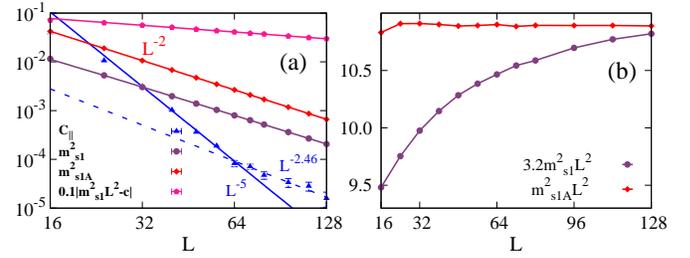}
\caption{(a) Log-log plot of the correlation function $C_\parallel(L/2)$, the squared magnetization $m^2_{\rm s1}$, the sublattice squared magnetization $m^2_{\rm s1A}$, and $|m^2_{\rm s1A}L^2-c|$ for the disordered phase, with $J^\prime_{\rm s}=2.5$ and $c=3.68$; (b) the data of  $m^2_{\rm s1}L^2$ and $m^2_{\rm s1A}L^2$.}
\label{jsp2.5}
\end{figure}

For the intermediate region $J^{\prime (1)}_{sc}<J^\prime_{s}<J^{\prime (2)}_{sc}$, the surface is in an ordinary phase.
As shown in Fig. \ref{jsp2.5}(a), the surface squared magnetization $m^2_{s1}$ shows a very good power-law scaling, which is the effect of the  nonsingular term $c/L^2$ because $y_{\rm h1}<1$ here.  This situation is very similar to the case of the $m^2_{\rm s1A}$ at the first special point .
In Fig. \ref{jsp2.5}(a), we also show the log-log plot of  $|m^2_{\rm s1}L^2-c|$ versus $L$, which also  indicates a good power-law scaling; the  data fitting according to Eq. (\ref{msurf_stag}) gives $y_{h1}=0.77(2)$, which is consistent with the result of the ordinary phase of the XY model. 

For the sublattice squared magnetization $m^2_{\rm s1A}$,  the nonsigular part is also dominated by the $L^{-2}$ scaling, as shown in Fig. \ref{jsp2.5}(a), however, we can not perform an effective fitting of $m^2_{\rm s1A}L^2$ according to  the scaling  formula (\ref{msA}), because $m^2_{\rm s1A}L^2$ converges to the constant $c$ too fast, as shown in Fig. \ref{jsp2.5}(b); it can be compared to the data of $m^2_{\rm s1}L^2$, which shows a very good asymptotic behavior, also shown in Fig. \ref{jsp2.5}(b). 

As to  the scaling behavior of the surface correlation function $C_\parallel(L/2)$,  
we find that in the short-range region, it decays very fast, with a power of about $L^{-5}$; however, in the long-range region, it decays slower with a power of about $L^{2y_{h1}-4}=L^{-2.46}$.  This is also illustrated in Fig. \ref{jsp2.5}(a). It should be noted that the dashed line in Fig. \ref{jsp2.5}(a) is not obtained by fitting the long-range part data of $C_\parallel(L/2)$ but only a self-consistent check with $y_{h1}=0.77$, which is got from the fitting of $m^2_{s1}$. 
The data in the long-range region is too small  which makes it too difficult to give accurate data in Monte Carlo simulations.

\subsection{Sublattice extraordinary-log phase}
\begin{figure}[htpb]
\includegraphics[width=0.8\columnwidth]{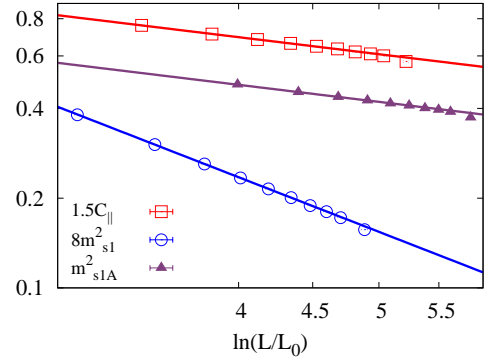}
\caption{Log-log plot of $C_{\parallel}(L/2)$, $m^2_{\rm s1}$,  and $m^2_{\rm s1A}$ versus $\ln(L/L_0)$, with $J^\prime_{\rm s}=3.8$ and $L_0=0.519$, 0.724, 0.295 for the three variables repectively. The slopes of the lines of $C_{\parallel}(L/2)$ and $m^2_{\rm s1A}$ are $q$=0.59, and the slope of $m^2_{\rm s1}$ is $q_1=1.9$.}
\label{jsp3.8}
\end{figure}
We also try to investigate the properties of the phase with $J^\prime_{\rm s}>J^{\prime (2)}_{\rm sc}$,
in this region, the efficiency of the algorithm decreases a lot, which prevents us to get good data for large systems. However, according to the already known results in our previous work\cite{Ding2022} that strong enough surface NN interactions will lead to an extraordinary-log phase, we infer that the phase should be an extraordinary-log one, furthermore, because the NNN interactions are already very strong, we infer that the extraordinary-log scaling may be dominated on the sublattices. To confirm such inference, we investigate the the scaling behaviors of the surface correlation function $C_\parallel(L/2)$ and 
the sublattice squared magnetization $m_{\rm s1A}^2$,  we find that they satisfy the logarithmic decaying formulas
\begin{eqnarray}
	C_\parallel(L/2)&=&a\cdot[\ln(L/L_0)]^{-q},\label{logCp}\\
	m^2_{\rm s1A}&=&a\cdot[\ln(L/L_0)]^{-q}.\label{logmA}
\end{eqnarray}
The value of the deacying exponent $q$ is the same for $C_\parallel(L/2)$ and $m^2_{\rm s1A}$,
which is $q=0.59(3)$, it is consistent with  that of the XY model.  An illustrative plot of $C_\parallel(L/2)$ and $m^2_{\rm s1A}$ is shown in Fig. \ref{jsp3.8}. 

We also investigate the squared magnetization $m_{\rm s1}^2$, we find that it also satisfy a logarithmical 
scaling formula
\begin{eqnarray}
	m^2_{\rm s1}=a\cdot[\ln(L/L_0)]^{-q_1}, \label{logm}
\end{eqnarray}
however, the  decaying exponent is found to be $q_1=1.9(2)$, which is much different from $q$; this result is also plotted in Fig. \ref{jsp3.8}. 

\begin{figure}[htpb]
	\includegraphics[width=1\columnwidth]{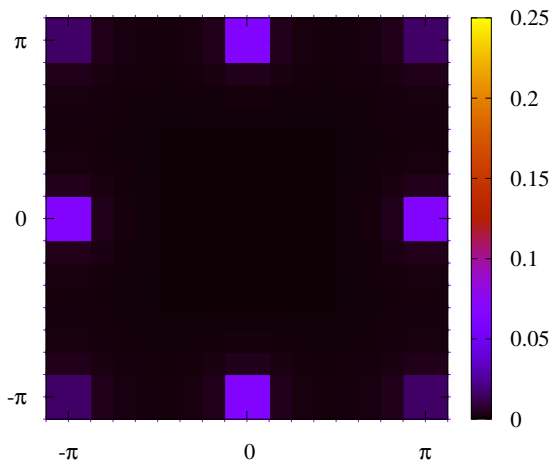}
	\caption{Surface structure factor $F(\vec{k})$ for the extraordinary phase, with $J^\prime_s=3.8$. The system size is $L=16$.}
	\label{fact}
\end{figure}

In order to further understand the properties of such sublattice extraordinary-log phase, we compute the surface structure factor
\begin{eqnarray}
F(\vec{k}))=\frac{1}{L^4}\Big\langle\big|\sum\limits_{\vec{R}}e^{i\vec{k} \cdot\vec{R}}\vec{\sigma}_{\vec{R}}\big|^2\Big\rangle   \label{F}
\end{eqnarray}
in the full momentum space ($k_x$, $k_y$).  
The result is shown in Fig. \ref{fact}, in which we can see that the points ($0,\pm\pi$) and ($\pm\pi,0$) are light, which means the system has a strip order (in finite system), this is in correspondence to the sublattice squared magnetization $m^2_{\rm s1A}$. The points ($\pm\pi,\pm\pi$) are also light but weaker than the points ($0,\pm\pi$) and ($\pm\pi,0$), this is in correspondence to the squared magnetization $m^2_{\rm s1}$.

\section{Conclusion and discussion}
\label{conclusion}
In summary, we have tried to add antiferromagnetic NNN interactions to the surface of the antiferromagnetic Potts model,  with the extraordinary-log phase as the starting point. We find that as the increasing of the NNN interactions, the system is driven to an ordinary phase in which the correlation function shares the same exponent of the ordinary phase of the XY model in the long-range region but decays much faster in the short-range region. Further strengthening the NNN interactions can  drive the surface to  a new extraordinary-log phase whose main properties are dominated on  the sublattices of the surface. The special transition from the disordered phase to the sublattice extraordinary-log phase belongs to a new universality class that is different to the well-known special transition of the XY model. 

It is well-known that the universality class of a phase transition is related to the broken symmetry if the phase transiton is between a disordered phase and a symmtery breaking phase.  In the surface criticality, it is shown that the universality calss of the special transition between a disordered phase (ordinary phase) and an extraordinary-log phase is also related to the symmetry of the model, although the extraordinary-log phase does not breaks the symmetry (without long-range order). 
For example, the special points of the classical O(2) and O(3) model belongs to different universality classes\cite{XYlog,Toldin2021}.   In the current paper, it is obvious that the new special point is also related to the symmetry of the sublattice extraordinary phase,  which not only has an O(2) symmetry of spins  but also a $Z_2$ symmetry of the lattice.


Surface criticality is a hybrid of the bulk criticality  and the surface physics.
In the study of the classical O($n$) model, different surface critical behaviors are obtained by tuning the surface NN interactions; especially,  the special point is a multicritical point that comes from the merge of the 3D criticality and the corresponding 2D criticality.
For an antiferromagnetic spin model, NNN interactions may lead to different universality class in 2D,  therefore, adding NNN interactions to the surface will lead to the interplay between the new 2D criticality and the 3D criticality, subsequently the surface criticality may be different.  This is a simple way to find new surface criticality; 
the special point between the extraordinary-log phase and the ordered phase presented in Ref. \onlinecite{Ding2022} and the new special point presented in the current paper are two examples of such purpose. Our results can be helpful in the  exploring of new surface critical behaviors.

\section*{Acknowledgment}
C.D. is supported by the National Science Foundation of China under Grants Number 11975024,
the Anhui Provincial Supporting Program for Excellent Young Talents in Colleges and Universities under Grant Number gxyqZD2019023.
L.Z. is supported by the National Key R\&D Program (2018YFA0305800), the National Natural Science Foundation of China under Grants Numbers 12174387 and 11804337,
CAS Strategic Priority Research Program (XDB28000000) and CAS Youth Innovation Promotion Association.
W.Z. was supported by the Hefei National Research Center for Physical Sciences at the Microscale (KF2021002).

\end{document}